\documentclass[12pt]{article}
\newcommand{\be}{\begin{equation}}
\newcommand{\ee}{\end{equation}}
\newcommand{\bea}{\begin{eqnarray}}
\newcommand{\eea}{\end{eqnarray}}
\newcommand{\ba}{\begin{array}}
\newcommand{\ea}{\end{array}}

\def\bbox{{\,\lower0.9pt\vbox{\hrule \hbox{\vrule height 0.2 cm
\hskip 0.2 cm \vrule height 0.2 cm}\hrule}\,}}
\newcommand{\dsl}{\pa \kern-0.5em /}

\newcommand{\call}{{\cal L}}

\begin{document}


\begin{titlepage}
\vfill
\begin{flushright}
\end{flushright}

\vfill
\begin{center}
\baselineskip=16pt
{\Large\bf Do Killing-Yano tensors form a Lie algebra?}
\vskip 1.0cm
{\large {\sl }}
\vskip 10.mm
{\bf David Kastor, Sourya Ray and Jennie Traschen} \\
\vskip 1cm
{

       Department of Physics\\
       University of Massachusetts\\
       Amherst, MA 01003\\
}
\vspace{6pt}
\end{center}
\vskip 0.5in
\par
\begin{center}
{\bf ABSTRACT}
\end{center}
\begin{quote}
Killing-Yano tensors are natural generalizations of Killing vectors.  We investigate whether Killing-Yano tensors form a graded Lie algebra with respect to the Schouten-Nijenhuis bracket.  We find that this proposition does not hold in general, but that it does hold for constant curvature spacetimes.  We also show that Minkowski and (anti)-deSitter spacetimes have the maximal number of Killing-Yano tensors of each rank and that the algebras of these tensors under the SN bracket are relatively simple extensions of the Poincare and (A)dS symmetry algebras.
\vfill
\vskip 2.mm
\end{quote}
\end{titlepage}
\section{Introduction}

The connection between symmetries and conservation laws in physics has both fundamental significance and great practical utility.    The symmetries of  curved spacetimes are generated by Killing vectors.   Killing vectors, in turn, yield conserved quantities,  both for geodesic motion within a given spacetime and also for the spacetime as a whole.   For geodesics, it is well known that if $K^a$ is a Killing vector and $u^a$ is the tangent vector to the geodesic (in affine parameterization), then the scalar  $u^a K_a$ is constant along the geodesic.   For the spacetime as a whole, the conserved Komar charge is given by $\int dS_{ab} \nabla^a K^b$, where the integral is taken over the boundary of a spacelike slice at infinity.  In addition,  ADM-like conserved charges may be defined for any class of spacetimes that are asymptotic at spatial infinity to a background spacetime admitting one or more Killing vectors \cite{Abbott:1981ff}.

Killing-Yano tensors are generalizations of Killing vectors that are also associated  with conserved quantities in a number of ways.  Let us call totally anti-symmetric contravariant tensor fields multivectors.  A Killing-Yano tensor of rank $n$ is a multivector $K^{a_1\dots a_n}$ that satisfies the property \cite{yano}
\begin{equation}\label{killing-yano}
\nabla_{(a_1} K_{a_2)a_3\dots a_{n+1}}=0.
\end{equation}
This clearly reduces to Killing's equation for $n=1$.  
We note that condition (\ref{killing-yano}) is equivalent to the statement
$\nabla_{a_1} K_{a_2a_3\dots a_{n+1}}=\nabla_{[a_1} K_{a_2a_3\dots a_{n+1}]}$.
Killing-Yano tensors of all ranks yield conserved quantities for geodesic motion.   
It is straightforward to show that the multivector $u^{a_1} K_{a_1}{}^{a_2 \dots a_n}$ is parallel transported alng a geodesic with tangent vector $u^a$.  
Such conservation laws, for example, underlie  the integrability of geodesics in the $d=4$ Kerr spacetime \cite{floyd}\cite{Penrose:1973um}
 and its higher dimensional generalizations \cite{Krtous:2006qy}\cite{Page:2006ka}\cite{Kubiznak:2006kt}.

It was further shown in reference 
\cite{Kastor:2004jk} that transverse asymptotically flat spacetimes  have conserved 
Y-ADM charges that are associated with the Killing-Yano tensors of flat spacetime.  Transverse asymptotcally flat spacetimes \cite{Townsend:2001rg}, such as those describing $p$-branes, become flat near infinity only in a subset of the full set of spatial directions, {\it i.e.} those directions 
transverse to the brane.
The construction of these Y-ADM charges, which are given by integrals over the boundary of  a codimension $n$ slice at transverse spatial infinity, follows closely the derivation of ADM conserved charges given in  \cite{Abbott:1981ff}.  
In a $p$-brane spacetime, the Y-ADM charges associated with rank $n=p+1$ Killing-Yano tensors are charge densities, {\it e.g.} mass per unit of world-volume area.  Positivity properties
of the Y-ADM mass density were studied in reference \cite{Kastor:2004tq}.  

Given that Killing-Yano tensors give rise to conserved quantities in general relativity, it is reasonable to ask whether Killing-Yano tensors are also associated with symmetries in some appropriately generalized sense.  In the case of  Killing vectors, the fact that they generate a group of continuous symmetries is reflected in their satisfying a closed Lie algebra. 
  One concrete way to investigate the possible association of Killing-Yano tensors with symmetries  is then to ask, as we will in this paper,  whether Killing-Yano tensors satisfy a closed Lie algebra amongst themselves?
  
  Part of the motivation for the present work is to lay ground work for developing a clearer understanding of the Y-ADM charges mentioned above.  For example,  in the Hamiltonian formalism, the ADM charges of asymptotically flat spacetimes satisfy a Poisson bracket algebra  that is isomorphic to the Lie algebra satisfied by the corresponding Killing vectors of Minkowski spacetime \cite{Regge:1974zd}, {\it i.e.} the Poincare algebra.  We would like to know if a similar structure holds for Y-ADM charges.   Because Y-ADM charges are given by  boundary integrals  on co-dimension $n$ submanifolds, we expect that the proper setting for such a bracket algebra will be a generalized Hamiltonian formalism in which data is similarly specified on a codimension $n$ submanifold and propagated via the Hamiltonian equations of motion in the remaining $n$ directions.  Such a formalism for general relativity has been developed in reference \cite{Grant:1997ai}.  We also need to know  whether the Killing-Yano tensors of flat spacetime satisfy a Lie algebra that generalizes the Poincare algebra of Killing vectors.   We will see below that, although Killing-Yano tensors do not generally form Lie algebras, that in fact they do in Minkowski or other maximally symmetric spacetimes.   Moreover, we will see that these Lie algebras are natural extensions of the Poincare and $(A)dS$ symmetry algebras.

Returning to Killing vectors, we know that the
 whole set of vector fields on a manifold $M$ form a Lie algebra. The Lie bracket of two vector fields $A^a$ and $B^a$ is defined to be the Lie derivative of $B^a$ with respect to $A^a$,
\begin{eqnarray}
[A,B] ^a &=&  ({\cal L}_A B){}^a\\
&=& A^b\partial_bB^a - B^b\partial_bA^a\label{liebracket}
\end{eqnarray}
This Lie algebra depends only on the manifold structure of $M$.  
The Killing vectors with respect to a particular spacetime metric $g_{ab}$ on $M$ turn out to form a subalgebra of this larger algebra of vector fields.  One way to see this is to consider Killing's equation in the form
\begin{equation}\label{killing}
({\cal L}_A g){}_{ab} = 0.
\end{equation}
The Lie derivative acting on tensor fields of arbitrary type satisfies the relation
\begin{equation}
{\cal L}_{[A,B]}=\call_A \call_B - \call_B\call_A .
\end{equation}
Therefore, if two vector fields $A^a$ and $B^a$ satisfy Killing's equation, so does the vector that results from taking their Lie bracket.
We would like to know whether a similar construction holds for Killing-Yano tensors.  Fortunately, 
it is known that the set of all multi-vector fields on a manifold $M$ form a (graded) Lie algebra with respect to the Schouten-Nijenhuis (SN) bracket \cite{Schouten1940, Schouten1954, Nijenhuis1955}. The SN bracket  serves as a starting point for our investigations below.

The plan of the paper is as follows.  In section (\ref{bracketintro}) we introduce the SN bracket and present some of its useful properties.  In section (\ref{calculations}) we study whether Killing-Yano tensors form a subalgebra of the full algebra of multi-vector fields with respect to the SN bracket.   We demonstrate that this property holds for constant curvature spacetimes.  However,  our calculations in section (\ref{calculations}) are  inconclusive as to whether it is true  in general spacetimes admitting Killing-Yano tensors.   In section (\ref{counterexample}) we look at  two examples of spacetimes, having non-constant curvature, that admit rank $2$ Killing-Yano tensors, $D=4$ Kerr and $D=4$ Taub-NUT.  In each of these cases, we find that the SN bracket of Killing-Yano tensors fails to satisfy the Killing-Yano equation (\ref{killing-yano}).   These serve as explicit counter-examples to the general conjecture.  In section (\ref{maxsym}) we present results on the Lie algebra of Killing-Yano tensors in maximally symmetric spaces.  We show that flat spacetime, as well as $(A)dS$ spacetimes admit the maximal number of Killing-Yano tensors of each rank and discuss the structures of the correcsponding Lie algebras.  We offer some concluding remarks in section (\ref{conclusions}).

\section{The Schouten-Nijenhuis Bracket}\label{bracketintro}

There exists  a  generalization to multivectors of the Lie bracket of vector fields that is  known as the Schouten-Nijenhuis (SN) bracket
\cite{Schouten1940,Schouten1954,Nijenhuis1955}.  
The SN bracket of  a  rank $p$ multivector $A$ and a rank $q$ multivector $B$ is a rank $(p+q-1)$ multivector, given in component form by 
\begin{eqnarray}\label{SNbracket}
[A,B]^{a_1\dots a_{p+q-1}} = p\, A^{b[a_1\dots a_{p-1}}
\partial_bB^{a_p\dots a_{p+q-1}]}
+ q\, (-1)^{pq} 
B^{b[a_1\dots a_{q-1}}
\partial_bA^{a_q\dots a_{p+q-1}]}
\end{eqnarray}
The spacetime dimension $D$ serves as the maximal rank for multivectors.  Therefore, the SN bracket vanishes if $p+q-1>D$.
The SN bracket reduces to the Lie bracket of vector fields (\ref{liebracket}) for $p=q=1$, and more generally for $p=1$ and $q$ arbitrary to the Lie derivative acting on a rank $q$ multivector.
Like the Lie bracket, the SN bracket is a map from tensors to tensors that depends only on the manifold structure.  The partial derivatives in the SN bracket may be replaced by covariant derivatives, as all the Christoffel symbol terms cancel out.  

The SN bracket is well known in the mathematics literature, playing a central role, for example, in the theory of Poisson manifolds.  The  properties of the SN bracket are discussed in detail in this conext in reference \cite{Vaisman}.  It is demonstrated there that the 
SN bracket  satisfies a number of important relations.  First is the exchange property
\begin{equation}\label{graded}
[A,B]=(-1)^{pq}[B,A]
\end{equation}
which shows the $Z_2$ grading of the SN bracket.  It can also be shown that the  SN bracket satisfies the graded product rule
\begin{equation}\label{gradedproduct}
[A,B\wedge C]=[A,B]\wedge C +(-1)^{(p+1)q}B\wedge[A,C]
\end{equation}
and the graded Jacobi identity
\begin{equation}\label{jacobi}
(-1)^{p(r+1)}[A, [B,C]]+ 
(-1)^{q(p+1)}[B, [C,A]] +
(-1)^{r(q+1)}[C, [A,B]]=0
\end{equation}
where $C$ is a multivector field of rank $r$.  
The wedge product between multivectors, denoted by the symbol $\wedge$ in equation (\ref{gradedproduct}), is defined in analogy with the wedge product of differential forms.
These relations imply that the set of multivector fields on a manifold satisfy a $Z_2$ graded Lie algebra.

For later use, we note  that the bracket of two  rank $2$ multivectors $A$ and $B$ is symmetric, and yields the 
rank $3$ multivector
\begin{equation}\label{twobracket}
[A,B]{}^{abc}=2( A^{d[a}\nabla_dB^{bc]} + B^{d[a}\nabla_dA^{bc]} ).
\end{equation}
One can easily check in this example that the Christoffel symbols on the right hand side cancel out.

\section{A Lie algebra of Killing-Yano tensors?}\label{calculations}

The next question to ask is whether the Killing-Yano tensors on a Riemannian manifold, {\it i.e.} multi-vector fields satisfying equation  (\ref{killing-yano}) with respect to the given metric, form a subalgebra of the full algebra of multivector fields.  Ideally at this point, we would want to proceed as we did with Killing vectors in the introduction, where we made use of Killing's equation in the form given in equation (\ref{killing}).    However, unlike the case of vector fields, the bracket of higher rank multivector fields cannot be extended to give an action of multivectors on tensors of general type analogous to the Lie derivative (see reference \cite{Vaisman} for a discussion of this point).
Thus, we have no way of rewriting the Killing-Yano condition (\ref{killing-yano}) in a form analogous to equation (\ref{killing}).

Instead we follow a more direct approach.  As a warm-up, we consider the Lie bracket of two Killing vectors and show that the resulting vector satisfies Killing's equation.  We then consider the SN bracket of a Killing vector with a rank $2$ Killing-Yano tensor and show that the resulting rank $2$ multivector is again a Killing-Yano tensor.  Finally, we consider the SN bracket of two rank $2$ Killing-Yano tensors and ask whether, or not, the resulting rank $3$ multivector satisfies the Killing-Yano condition (\ref{killing-yano}).  Our result, in this case, is inconclusive.  We go on to show that for constant curvature spacetimes, that the answer is yes.   However, in section (\ref{counterexample}) we give two explicit counterexamples which demonstrate that the property does not hold in general.

Before proceeding with our calculations, it is worth noting that there is also an SN bracket for totally symmetric contravariant tensors, which gives the set of all such fields on a manifold a natural Lie algebra structure.  The symmetric SN bracket may be used to show that the set of symmetric Killing tensors form a  Lie subalgebra of this larger algebra of all symmetric tensor fields.  The proof is very similar to the one given for Killing vectors in the introduction.  Killing tensors are defined by the condition 
\begin{equation}\label{killingtensor}
\nabla_{(a_1}K_{a_2\dots a_n)}=0.
\end{equation}
The inverse metric $g^{ab}$, in particular, is clearly a symmetric Killing tensor.  
One can show that equation
(\ref{killingtensor}) is equivalent to the condition that the SN bracket of $K^{a_1\dots a_n}$ with the 
inverse metric vanishes \cite{geroch}.  It then follows from the Jacobi identity for the symmetric SN bracket that the bracket of two symmetric Killing tensors is again a symmetric Killing tensor\footnote{The SN bracket plays a natural role in the Hamiltonian formulation of geodesic motion. If 
$K^{a_1\dots a_n}$ is a symmetric Killing tensor, then the scalar quantity 
$K^{a_1\dots a_n}u_{a_1}\dots u_{a_n}$ is conserved along a geodesic with tangent vector $u^a$.  
The condition that the Poisson bracket of two such conserved quantities vanishes is simply that the SN bracket of the two Killing tensors vanishes.  See reference \cite{thompson} for a discussion of this topic.}.
Killing vectors may equally well be considered to be either totally symmetric, or totally anti-symmetric.  It seems interesting that with regard to the considerations above, Killing vectors seem to have more in common with symmetric Killing tensors.

Returning to Killing-Yano tensors, we proceed in a direct manner to check whether, or not,  the SN bracket preserves the Killing-Yano property.   It is instructive to  go through the calculation first for Killing vectors.  The necessary ingredient in this calculation is the property that acting with two derivatives on a Killing vector $A^a$ gives
\begin{equation}\label{twoderivatives}
\nabla_a\nabla_bA_c = - R_{bca}{}^dA_d.
\end{equation}
Let $A^a$ and $B^a$ be two Killing vectors.   In order to check that their bracket $C^a=A^b\nabla_bB^a-B^b\nabla_bA^a$ also satisfies Killing's equation, we compute
\begin{eqnarray}
\nabla_aC_b &=& (\nabla_a A_c)\nabla^cB_b - (\nabla_a B_c)\nabla^cA_b
+A_c\nabla_a\nabla^c B_b - B_c\nabla_a\nabla^c A_b \\
&=& -(\nabla_c A_a)\nabla^c B_b + (\nabla_c A_b)\nabla^c B_a 
-R_{abcd}A^cB^d\\
&=& \nabla_{[a}C_{b]},
\end{eqnarray}
where Killing's equation for $A^a$ and $B^a$  is used in processing the first derivative terms and
equation (\ref{twoderivatives}) is used together with basic properties of the Riemann tensor in processing the second derivative terms.  

We now proceed to show that the SN bracket of a Killing vector $A^a$ with a Killing-Yano tensor $B^{ab}$ is again a Killing-Yano tensor.  Let $C^{ab}=[A,B]^{ab}$, which in this case is simply the Lie derivative of $B^{ab}$ with respect to the Killing vector $A^a$.  The calculation requires the analogue of equation (\ref{twoderivatives})   for a rank $2$ Killing-Yano tensor.  This  is given by
\begin{equation}\label{kytidentity}
\nabla_a\nabla_b B_{cd}= {3\over 2} R_{[bc|a|}{}^e\, B_{d]e}
\end{equation}
The starting point is then the expression 
$C^{ab}=A^c\nabla_c B^{ab}- B^{cb}\nabla_cA^a-B^{ac}\nabla_c A^b$.  
Making use of the Killing-Yano condition for $A^{ab}$ and $B^{ab}$, equation (\ref{kytidentity}) and properties of the Riemann tensor,
one can show that
\begin{equation}
\nabla_aC_{bc} = -3(\nabla_dA_{[a})\nabla^dB_{bc]}  
+{3\over 2}A^d R_{de[ab}B_{c]}{}^e
\end{equation}
We see that $\nabla_aC_{bc} =\nabla_{[a}C_{bc]} $ and that therefore $C_{ab}$ is again a Killing-Yano tensor.

Finally, let us now assume that $A^{ab}$ and $B^{ab}$ are two rank $2$ Killing-Yano tensors and ask whether, or not, $C^{abc}=[A,B]^{abc}$ is also a Killing-Yano tensor? From equation (\ref{twobracket}), we have that 
\begin{equation}
C^{abc}= 2( A^{d[a}\nabla_dB^{bc]} + B^{d[a}\nabla_dA^{bc]} ).
\end{equation}
In this case, we were not able to show generally that $\nabla_a C_{bcd}= \nabla_{[a} C_{bcd]}$.
There are many equivalent expressions that may be given for $\nabla_a C_{bcd}$, one of which is
\begin{equation}
\nabla_aC_{bcd}  =  - 4 (\nabla_e A_{[ab})\nabla^eB_{cd]} +
2\left( A_{e[b}\nabla_{|a|}\nabla^eB_{cd]} + B_{e[b}\nabla_{|a|}\nabla^eA_{cd]} \right). \label{doesitwork}
\end{equation}
Here we see that the terms quadratic in first derivatives, which have been processed using the Killing-Yano condition $A^{ab}$ and $B^{ab}$, do display the necessary anti-symmetry consistent with $C^{abc}$ having the Killing-Yano property.   
The second derviative terms may be rewritten in many equivalent forms using equation (\ref{kytidentity}) and the identities 
$\nabla_{[a}\nabla_b A_{cd]}=\nabla_{[a}\nabla_b B_{cd]}=0$.
However, none of these forms appears to be totally anti-symmetric in the free indices.  We confirm the conclusion that $C^{abc}$ is not generally a Killing-Yano tensor in the next section, by presenting two 
explicit counterexamples.

Our conclusion, however, does not imply that the SN bracket of Killing-Yano tensors is never a Killng-Yano tensor.  It is interesting to consider the special case of constant curvature spacetimes.  Constant curvature spacetimes locally have the maximal number of Killing-Yano tensors of each rank.  Specializing even further to flat spacetime, the second derivative terms in (\ref{doesitwork}) vanish by virtue of equation (\ref{kytidentity}).  Hence, given the anti-symmetry of the remaining term on the right hand side of (\ref{doesitwork}), we see that the Killing-Yano tensors of flat spacetime do form a graded Lie algebra with respect to the SN bracket.  More generally, for a constant curvature spacetime the Riemann tensor is given by $R_{abcd}=\alpha(g_{ac}g_{bd}-g_{bc}g_{ad})$ and we find that a rank $2$ Killing-Yano tensor satisfies
\begin{equation}
\nabla_a\nabla_b A_{cd}= - 3\alpha g_{a[b}A_{cd]}.
\end{equation}
Equation (\ref{doesitwork}) then becomes
\begin{equation}\label{ranktwo}
\nabla_aC_{bcd} =  - 4 (\nabla_e A_{[ab})\nabla^eB_{cd]} - 4\alpha A_{[ab}B_{cd]} = \nabla_{[a}C_{bcd]}
\end{equation}
and we see that the SN bracket of rank $2$ Killing-Yano tensors in a constant curvature spacetime is again a Killing-Yano tensor.

It is straightforward to show that the SN bracket always preserves the Killing-Yano property in spacetimes with constant curvature.  Acting with two derivatives on a Killing-Yano tensor of rank $n$ gives
\begin{equation}
\nabla_a\nabla_{b_1} A_{b_2\dots b_{n+1}}= (-1)^{n+1}\left({n+1\over 2}\right)
R^d{}_{a[b_1b_2}A_{b_3\dots b_{n+1}]d}
\end{equation}
For constant curvature spacetimes, we then have
\begin{equation}
\nabla_a\nabla_{b_1} A_{b_2\dots b_{n+1}}=-(n+1)\alpha g_{a[b_1}A_{b_2\dots b_{n+1}]}.
\end{equation}
Let $A$ and $B$ be Killing-Yano tensors of ranks $m$ and $n$ respectively, then $C=[A,B]$ is a multivector of rank $m+n-1$.  One finds that 
\begin{eqnarray}
\nabla_{a_1}C_{a_2\dots a_{m+n}} = &&  -(m+n)  \left\{
(\nabla_cA_{[a_1\dots a_m})\nabla^c B_{a_{m+1}\dots a_{m+n}]} \right .\\ \nonumber  &&  + \left .
\alpha A_{[a_1\dots a_m}B_{a_{m+1}\dots a_{m+n}]}  \right\},
\end{eqnarray}
which reduces to equation (\ref{ranktwo}) for $m=n=2$ and shows generally that the Killing-Yano tensors of a constant curvature spacetime form a graded Lie algebra under the SN bracket.

\section{Two counterexamples}\label{counterexample}

In the preceding section, we were unable to verify, or definitively falsify, the proposition that the  SN bracket of two Killing-Yano tensors is necessarily another Killing-Yano tensor, for the general case of metrics with non-constant curvature.  
In this section, we show that the  proposition is indeed false by providing two explicit counterexamples\footnote{The proposition that the SN bracket preserves the Killing-Yano property has been previously investigated in reference \cite{dolan}.  It is claimed there, without a proof being presented, that the propositions is valid.  The counter-examples presented in this section demonstrate that this is not the case.}.  

\subsubsection*{Counterexample \# $1$: $D=4$ Kerr}
The first counterexample is in
the $4$-dimensional Kerr spacetime, which has a rank $2$ Killing-Yano tensor $A^{ab}$, which was originally found by Penrose \cite{Penrose:1973um} and Floyd \cite{floyd}.   The tensor $A^{ab}$  plays a central role in the integrability of geodesic motion and the separability of various wave equations (see reference \cite{Gibbons:1993ap} for a discussion and further references).  Recall from equation (\ref{graded}) that the SN bracket is  $Z_2$ graded, and in particular that the bracket of two rank $2$ multivectors is symmetric under interchange.  Therefore,  one can consider the tensor  $B^{abc}=[A,A]^{abc}$ and ask whether, or not, it is a Killing-Yano tensor?

The Kerr metric in Boyer-Lindquist coordinates is given by
\begin{equation}
ds^2 = -{\Delta\over\rho^2}(dt-a\sin^2\theta d\phi)^2 + {\sin^2\theta\over\rho^2}((r^2+a^2)d\phi-adt)^2 
+ {\rho^2\over\Delta}dr^2 +\rho^2 d\theta^2,
\end{equation}
where $\rho^2=r^2+a^2\cos^2\theta$ and $\Delta=r^2+a^2-2mr$.
The nonzero components of the Killing-Yano tensor $A^{ab}$ in these coordinates are (from reference \cite{Gibbons:1993ap})
\begin{eqnarray}
A^{rt}&=& -{a\cos\theta(r^2+a^2)\over \rho^2}, \qquad A^{r\phi}=-{a^2\cos\theta\over\rho^2},\qquad
\\ \nonumber
A^{\theta t}&=&{ar\sin\theta\over\rho^2},\qquad A^{\theta\phi}={r\over\rho^2\sin\theta}
\end{eqnarray}
The  tensor  $B^{abc}=[A,A]^{abc}$ may then be computed from equation (\ref{twobracket}), giving the nonzero components
\begin{equation}
B^{rt\phi}={4ar\over 3\rho^2},\qquad B^{\theta t\phi} = {4a\cos\theta\over 3\rho^2\sin\theta}.
\end{equation}

We now want to check whether, or not, the tensor $B^{abc}$ is itself a Killing-Yano tensor.
It requires very little work to check that the condition $\nabla_aB^{abc}=0$ is satisfied.  This is consistent with the Killing-Yano property (\ref{killing-yano}).  However, checking in more detail whether equation (\ref{killing-yano}) is satisfied, it turns out that many components of the tensor $\nabla_{(a}B_{b)cd}$ are non-vanishing.  For example, one finds that
\begin{equation}\label{notworking}
\nabla_rB_{\theta t\phi}+\nabla_\theta B_{rt\phi}= {2\over 3}\; am \sin 2\theta
\left({a^2\cos^2\theta-3r^2\over \rho^2}\right)
\end{equation}
We see, therefore, that the SN bracket of the Killing-Yano tensor $A^{ab}$ of $D=4$ Kerr with itself does not satisfy the Killing-Yano property.  
The only exception to this is the case $m=0$ (note that the tensor $B^{abc}$ vanishes for $a=0$).  In this case, all the components of $\nabla_{(a}B_{b)cd}$ vanish and hence $B^{abc}$ is a Killing-Yano tensor.  
This is consistent with our results of section (\ref{calculations}).
For $m=0$, the Kerr metric is flat, and we have shown in section  (\ref{calculations}) that the SN bracket of Killing-Yano tensors is always a Killing-Yano tensor in flat spacetime.  The counterexample presented above, however, serves to falsify the general proposition that the SN bracket of Killing-Yano tensors is always a Killing-Yano tensor.

\subsubsection*{Counterexample \# $2$:  Euclidean Taub-NUT}

We have now seen the SN brackets of Killing-Yano tensors are not generally Killing-Yano tensors.  However, we have also seen in section (\ref{calculations}) that this property does hold in the special case of constant curvature.  It would be interesting to know whether there are other spacetimes for which the property holds as well.  The result for constant curvature spacetimes suggests that we look at 
other spacetimes with Killing-Yano tensors for which the Riemann tensor enjoys some other special property.  The Euclidean Taub-NUT metric has a self-dual Riemann tensor.  It admits four rank $2$ Killing-Yano tensors  (see reference \cite{Gibbons:1987sp} for a detailed discussion).  We have checked whether the SN brackets of these tensors satisfy the Killing-Yano condition and found that they do not.

The metric of the Euclidean Taub-NUT spacetime is given by
\begin{equation}
ds^2 = V(r)(dr^2+r^2 d\theta^2+r^2\sin^2\theta d\phi^2) +  {16 m^2\over V(r)}(d\chi +\cos\theta d\phi)^2,
\quad V(r) = 1 + {4m\over r}
\end{equation}
The four rank $2$ Killing-Yano tensors are given in covariant form by
\begin{eqnarray}
f_Y &=& 8m(d\chi+\cos\theta d\phi)\wedge dr + 4r(r+2m)(1+{r\over 4m})\sin\theta d\theta\wedge d\phi \\
f_i &=& 8m(d\chi+\cos\theta d\phi)\wedge dx_i -\epsilon_{ijk} dx_i\wedge dx_j
\end{eqnarray}
where $i,j=1,2,3$ and the Cartesian coordinates $x_i$ are related to the spherical coordinates $(r,\theta,\phi)$ in the standard way.  The $2$-forms $f_i$ are covariantly constant and therefore have trivially vanishing SN brackets.  The tensors $B^{abc}=[f_Y,f_Y]^{abc}$ and $B^{(i)abc}=[f_i,f_Y]^{abc}$
are nonzero.  However, we find that the tensors $\nabla_a B_{bcd}$ and 
$\nabla_a B^{(i)}_{bcd}$ are not totally anti-symmetric.  Therefore $B^{abc}$ and $B^{(i)abc}$ are not Killing-Yano tensors.

\section{Maximally symmetric spacetimes}\label{maxsym}

It was shown in section (\ref{calculations}) that the Killing-Yano tensors of a constant curvature spacetime do form a Lie algebra with respect to the SN bracket.
In this section, we will study the Killing-Yano tensors of $d$-dimensional Minkowski and (anti-)deSitter spacetimes.  These spacetimes are well known to have the maximal number of Killing vectors.  We will see that they also have the maximal number of Killing-Yano tensors\footnote{Y-ADM charges for asymptotically $AdS$ spacetimes have been studied in \cite{Cebeci:2006mc}\cite{ads-unpublished}.}. 
Calculation of the full Lie algebra of Killing-Yano tensors proves to be quite tedious.  It will be clear, however, from the explicit forms of the Killing-Yano tensors that these algebras are, in principle,  simple extensions of the Poincare and $(A)dS$ algebras.

Let us denote the maximal number of Killing-Yano tensors of rank $n$ in a given spacetime dimension $d$ by $N(d,n)$.  This number is determined  \cite{Kastor:2004jk} via a simple generalization of the argument for Killing vectors.  By virtue of equation (\ref{twoderivatives}), a Killing vector $A^a$ is determined everywhere on a manifold by its value at one point together with the values of its first derivatives $\nabla_a\xi_b$ at that point (see  {\it e.g.} appendix C of reference \cite{Wald:rg}).  Since $\nabla_a\xi_b$ is antisymmetric, it follows that there are at most $N(d,1)=d+d(d-1)/2$ linearly independent Killing vectors.

The counting works similarly for Killing-Yano tensors of rank $n$, where a generalization of equation 
(\ref{kytidentity}) determines all second and higher order derivatives in terms of the values of the tensor and its first derivatives.  Since  Killing-Yano tensor $A_{a_1\dots a_n}$ and its  first derivative $\nabla_{a_1}A_{a_2\dots a_{n+1}}$  are both anti-symmetric, it follows that
\begin{equation}
N(d,n)= {d!\over n! (d-n)!} + {d!\over (n+1)! (d-(n+1))!} 
\end{equation}

Let us start with Minkowski spacetime and denote the $d$ translational Killing vectors by $T_{<a>}$, where the label $a$ runs over the $d$ spacetime coordinates.  In component form, we have simply $T_{<a>}^b=\delta_a^b$.  Similarly, we will denote the $d(d-1)/2$ boost and rotation Killing vectors by
$R_{<ab>}$, with $R_{<ba>}=-R_{<ab>}$.  These have component form 
$R_{<ab>}^c=2x^d\eta_{d[a}\delta_{b]}^c$.  The higher rank Killing-Yano tensors may similarly be classified as either translations or boost/rotations.  The number of translational and boost/rotational Killing-Yano tensors of rank $n$ are given respectively by
\begin{equation}
N_T(d,n)= {d!\over n! (d-n)!},\qquad N_R(d,n)={d!\over (n+1)! (d-(n+1))!} 
\end{equation}
which together add up to the maximal number.   The rank $n$ translations and boost/rotations can be labeled $T_{<a_1\dots a_n>}$ and $R_{<a_1\dots a_{n+1}>}$.  They are antisymmetric in their label indices.  Their components forms are
\begin{eqnarray}
&&T_{<a_1\dots a_n>}^{b_1\dots b_n}= n! \delta_{[a_1}^{b_1}\cdots\delta_{a_n]}^{b_n}\\
&&R_{<a_1\dots a_{n+1}>}^{b_1\dots b_{n}}= (n+1)! x^c \eta_{c[a_1}\delta_{a_2}^{b_1}\cdots\delta_{a_{n+1}]}^{b_n}
\end{eqnarray}
It is straightforward to check that these tensors satisfy the Killing-Yano condition (\ref{killing-yano}).
It is likewise straightforward, in principle, to calculate the SN brackets of the collection of whole complex of Killing-Yano tensors of Minkowski spacetime.  However, the many anti-symmetrizations make this a tedious task.  It is simple, however, to see the general form that the full algebra of Killing-Yano tensors take.  Translational Killing-Yano tensors are independent of the spacetime coordinates, while boost/rotations are linear in the spacetime coordinates.  The SN bracket has one derivative.  Therefore, the full algebra of Killing-Yano tensors will have a structure very similar to that of the Poincare algebra.  The bracket of two translations will vanish.  The bracket of a translation with a boost/rotation will be a translation, while the bracket of two boost/rotations will be another boost/rotation.  Hence, the boosts and rotations form a Lorentz-like subalgebra.

We now turn to $(A)dS_d$, which we
realize  as  hyperboloids in $(d+1)$ dimensional flat spacetime with coordinates $X^A$ with $A=0,\dots,d$  and signature $(d,1)$ for $dS_d$ and $(d-1,2)$ for $AdS_d$.  The hyperboloids are then 
given by 
\begin{equation}\label{hyperboloid}
-(X^0)^2 +(X^1)^2 +\dots (X^{d-1})^2 +\kappa (X^d)^2 = R^2,
\end{equation}
where $\kappa=+1$ for $dS_d$ and $\kappa=-1$ for $AdS_d$.  If we now introduce coordinates $x^a$ with $a=0,\dots,d-1$ on the $(A)dS_d$ hyperboloids and use the radius $R$ as an additional coordinate, then the flat $(d+1)$ dimensional metric may be written as
\begin{equation}
ds^2_{n+1}= \kappa dR^2 +R^2 k_{ab}dx^a dx^b,
\end{equation}
where $g_{ab}=R^2 k_{ab}$ is the $(A)dS_n$ metric and $k_{ab}=k_{ab}(x^c)$.   Let $\nabla_a$ denote the $(A)dS_d$ covariant derivative operator and $\xi^a$ the projection of a flat space Killing vector  onto the $(A)dS_d$ hyperboloid.  One can then show that 
\begin{equation}\label{extrinsic}
\nabla_a \xi_b + \nabla_b \xi_a = - {2\kappa\over R} g_{ab} \xi_R,
\end{equation}
where $\xi^R$ is the component of the Killing vector field normal to the hyperboloid.  Killing vectors of the flat embedding space that are tangent to the hyperboloid are then seen to be Killing vectors of $(A)dS_d$.  The rotational Killing vectors of the flat embedding spacetime satisfy this property.  There are $N_R(d+1,1)$ of these vectors, and since $N_R(d+1,1)=N(d,1)$ we see that $(A)dS_d$ are maximally symmetric spacetimes.   We can also see from equation (\ref{extrinsic}) that the projections of the translational Killing vectors of the flat embedding space onto the $(A)dS_d$ hyperboloid yield conformal Killing vectors, since for these $\xi^R$ will be non-zero.

The situation for higher rank Killing-Yano tensors is quite similar.  Let $\xi^{a_1\dots a_n}$ be the projection of a rank $n$ Killing-Yano tensor of the flat embedding spacetime onto its components tangent to the $(A)dS_d$ hyperboloid.  The tensor $\xi^{a_1\dots a_n}$ then satisfies the equation
\begin{eqnarray}\label{extrinsicyano}
2 \nabla_{(a_1} \xi_{a_2)a_3\dots a_{n+1}}  & = &
- { \kappa\over R} \{ g_{a_1 a_2} \xi_{Ra_3\dots a_{n+1} }     + 
 \dots + g_{a_1 a_{n+1} } \xi_{a_2a_3\dots a_r R }  \\ \nonumber &&
 + g_{a_2 a_1}  \xi_{Ra_3\dots a_{n+1} } +\dots + g_{a_2 a_{n+1} } \xi_{a_1a_3\dots a_r R }\}.
\end{eqnarray}
Killing-Yano tensors of the flat embedding spacetime that are tangent to the hyperboloid, {\it i.e.} for which $\xi_{Ra_1\dots a_{n-1} }=0$, are then also Killing-Yano tensors of $(A)dS_d$.  It is straightforward to check that the rotational Killing-Yano tensors $R_{<a_1\dots a_{n+1}>}$  have this property.  Moreover,  they provide precisely the maximal number of $(A)dS_d$ Killing-Yano tensors of each rank.  This can be seen by noting the equality
\begin{equation}\label{comb}
N(d,n) =N_R(d+1,n)= {(d+1)!\over n! (d+1-n)!}
\end{equation}
Hence $(A)dS_d$ has the maximal number of Killing-Yano tensors of each rank.
We also see from this construction that the Lie algebra of Killing-Yano tensors for $(A)dS_d$ coincides with the subalgebra of boost/rotational Killing-Yano tensors in the corresponding flat $(d+1)$-dimensional embedding space.

The translational Killing-Yano tensors of the embedding spacetime have non-vanishing radial components.  Their projections onto the $(A)dS_d$ hyperboloid yield conformal Killing-Yano tensors.  These are defined in reference \cite{Kashiwada} to be antisymmetric tensor fields satisfying the condition
\begin{eqnarray}
 \nabla_b \xi_{a_1\dots a_p} + \nabla _{a_1}\xi_{ba_2\dots a_p} &=&
2g_{ba_1}\chi_{a_2\dots a_p} - \\ \nonumber &&
\sum_{i=2}^p (-1)^i (g_{ba_i}\chi_{a_1\dots \hat a_i \dots a_p} +
g_{a_1a_i}\chi_{ba_2\dots \hat a_i \dots a_p})
\end{eqnarray}
for some antisymmetric tensor $\chi_{a_1\dots a_{p-1}}$ of one degree lower rank.

\section{Conclusions}\label{conclusions}

We have shown that, although the proposition that the SN bracket preserves the Killing-Yano property is false in general, it does hold at least locally for constant curvature spacetimes.  Through a counting argument in section (\ref{maxsym}), we have also found the maximum possible number of Killing-Yano tensors of a given rank in a given spacetime dimension.  
We have then seen via explicit construction that Minkowski and $(A)dS$ spacetimes have the maximal possible number of Killing-Yano tensors of each rank.  The algebra of these tensors under the SN bracket extends the structure of the Poincare and $(A)dS$ Lie algebras respectively.

These results suggest a number of interesting questions for additional work.  A few of these are the following.  Are there other classes of spacetimes, in which the Killing-Yano tensors form an algebra under the SN bracket?  Do the Y-ADM charges of asymptotically flat and $(A)dS$ spacetimes, as discussed in the introduction,  form algebras with respect to some form of generalized Poisson brackets?  Finally, we can ask whether the representation theories of the graded Lie algebras of Killing-Yano tensors for Minkowski and $(A)dS$ spacetimes have any physical significance?

\subsection*{Acknowledgements}
We thank Ivan Mirkovic for helpful conversations.  This work was supported by NSF grant PHY-0555304

\end{document}